\documentclass[fleqn,100pt]{wlscirep}
\usepackage{amsmath,amssymb,wasysym}
\usepackage{xcolor}

\usepackage{siunitx}
\usepackage{lineno}
\title{Anisotropic magnetocaloric effect in Fe$_{3-x}$GeTe$_2$}
\author[1*]{Yu Liu}
\author[1]{Jun Li}
\author[1]{Jing Tao}
\author[1]{Yimei Zhu}
\author[1*]{Cedomir Petrovic}
\affil[1]{Condensed Matter Physics and Materials Science Department, Brookhaven National Laboratory, Upton, New York 11973, USA}
\affil[*]{yuliu@bnl.gov and petrovic@bnl.gov}

\begin{abstract}
We present a comprehensive study on anisotropic magnetocaloric porperties of the van der Waals weak-itinerant ferromagnet Fe$_{3-x}$GeTe$_2$ that features gate-tunable room-temperature ferromagnetism in few-layer device. Intrinsic magnetocrystalline anisotropy is observed to be temperature-dependent and most likely favors the long-range magnetic order in thin Fe$_{3-x}$GeTe$_2$ crsytal. The magnetic entropy change $-\Delta S_M$ also reveals an anisotropic characteristic between $H // ab$ and $H // c$, which could be well scaled into a universal curve. The peak value $-\Delta S_M^{max}$ of 1.20 J kg$^{-1}$ K$^{-1}$ and the corresponding adiabatic temperature change $\Delta T_{ad}$ of 0.66 K are deduced from heat capacity with out-of-plane field change of 5 T. By fitting of the field-dependent parameters of $-\Delta S_M^{max}$ and the relative cooling power RCP, it gives $-\Delta S_M^{max} \propto H^n$ with $n = 0.603(6)$ and $RCP \propto H^m$ with $m = 1.20(1)$ when $H // c$. Given the high and tunable $T_c$, Fe$_{3-x}$GeTe$_2$ crystals are of interest for fabricating the heterostructure-based spintronics device.
\end{abstract}
\begin{document}

\flushbottom
\maketitle
% * <john.hammersley@gmail.com> 2015-02-09T12:07:31.197Z:
%
%  Click the title above to edit the author information and abstract
%
\thispagestyle{empty}

\section*{Introduction}

Intrinsic long-range ferromagnetism recently achieved in two-dimensional-limit van der Waals (vdW) crystals opens up great possibilities for both studying fundamental two-dimensional (2D) magnetism and engineering novel spintronic vdW heterostuctures.\cite{McGuire,Huang,Seyler,Gong,Fei} Fe$_3$GeTe$_2$ is a promising candidate since its Curie temperature ($T_c$) in bulk is high and depends on the concentration of Fe atoms, ranging from 150 to 230 K.\cite{Deiseroth,Chen,May,Liu,Yi,BJLiu} Intrinsic magnetocrystalline anisotropy in few-layer counteracts thermal fluctuation and favors the 2D long-range ferromagnetism with a lower $T_c$ of 130 K.\cite{Fei} Most significantly, the $T_c$ can be ionic-gate-tuned to room temperature in few-layers which is of high interest for electrically controlled magnetoelectronic devices.\cite{Deng}

The layered Fe$_{3-x}$GeTe$_2$ displays a hexagonal structure belonging to the P6$_3$/mmc space group, where the 2D layers of Fe$_{3-x}$Ge sandwiched between nets of Te ions are weakly connected by vdW bonding [Fig. 1(a)].\cite{Deiseroth} There are two inequivalent Wyckoff positions of Fe atoms which are denoted as Fe1 and Fe2. The Fe1-Fe1 dumbbells are situated in the centre of the hexagonal cell in the honeycomb lattice, composed of covalently bonded Fe2-Ge atoms. No Fe atoms occupy the interlayer space and Fe vacancies only occur in the Fe2 sites.\cite{VerchenkoVY} Local atomic environment is also studied by the M\"{o}ssbauer and X-ray absorption spectroscopies.\cite{YULIU,YULIU1} Partially filled Fe $d$ orbitals results in an itinerant ferromagnetism in Fe$_{3-x}$GeTe$_2$,\cite{Zhuang} which exhibits exotic physical phenomena such as nontrivial anomalous Hall effect,\cite{Kim,Wang,Tan} Kondo lattice behavior,\cite{Zhang} strong electron correlations,\cite{Zhu} and unusual magnetic domain structures.\cite{Nguyen,Leon} A second-step satellite transition $T^*$ is also observed just below $T_c$, and is not fully understood.\cite{Yi,YULIU1}

Here we address the anisotropy in Fe$_{3-x}$GeTe$_2$ as well as the magnetocaloric effect investigated by heat capacity and dc magnetization measurements. The magnetocrystalline anisotropy is observed to be temperature-dependent. The magnetic entropy change $\Delta S_M(T,H)$ also reveals an anisotropic characteristic and could be well scaled into a universal curve. Moreover, the $-\Delta S_M^{max}$ follows the power law of $H^n$ with $n = 0.603(6)$, and the relative cooling power RCP depends on $H^m$ with $m = 1.20(1)$.

\section*{Methods}

High quality Fe$_{3-x}$GeTe$_2$ single crystals were synthesized by the self-flux technique.\cite{YULIU} The element analysis was performed using energy-dispersive X-ray spectroscopy (EDX) in a JEOL LSM-6500 scanning electron microscope (SEM). The selected area electron diffraction pattern was taken via a double aberration-corrected JEOL-ARM200F operated at 200 kV. The dc magnetization and heat capacity were measured in Quantum Design MPMS-XL5 and PPMS-9 systems with the field up to 5 T.

\section*{Results and Discussion}

\begin{figure*}
\centerline{\includegraphics[scale=0.9]{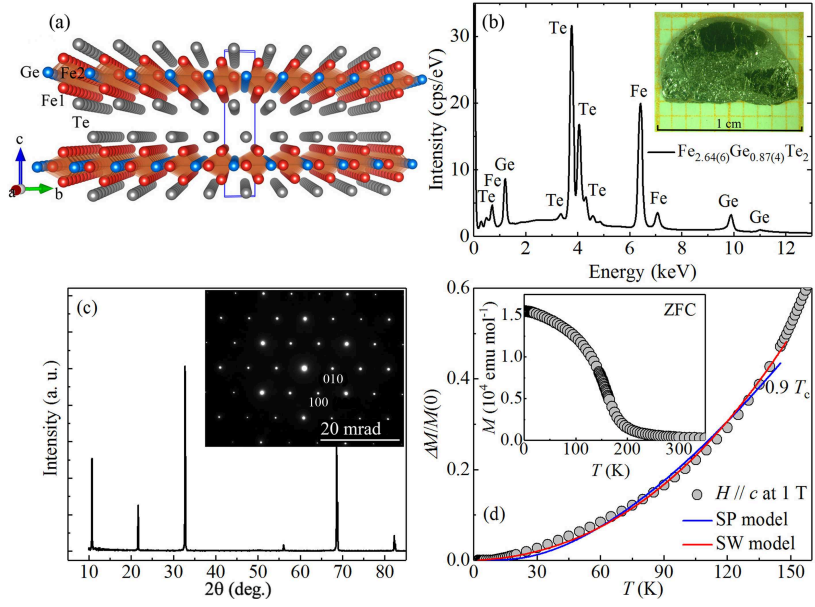}}
\caption{(Color online). (a) Crystal structure and (b) X-ray energy-dispersive spectrum of Fe$_{3-x}$GeTe$_2$ single crystal. Inset shows a photograph of Fe$_{3-x}$GeTe$_2$ single crystal on a 1 mm grid. (c) X-ray diffraction (XRD) pattern of Fe$_{3-x}$GeTe$_2$. Inset shows the electron diffraction pattern taken along the [001] zone axis direction. (d) Temperature dependence of the reduced magnetization with out-of-plane field of Fe$_{3-x}$GeTe$_2$ fitted using spin-wave (SW) model and single-particle (SP) model. Inset shows the temperature dependence of zero-field-cooling (ZFC) magnetization of Fe$_{3-x}$GeTe$_2$ measured at $H$ = 1 T applied along the $c$ axis.}
\label{1}
\end{figure*}

The average stoichiometry of our flux-grown Fe$_{3-x}$GeTe$_2$ single crystals was determined by examination of multiple points. The actual concentration is determined to be Fe$_{2.64(6)}$Ge$_{0.87(4)}$Te$_2$ [Fig. 1(b)], and it is referred to as Fe$_{3-x}$GeTe$_2$ throughout this paper. The as-grown single crystals are mirror-like and metallic platelets with the crystallographic $c$ axis perpendicular to the platelet surface with dimensions up to 10 millimeters [inset in Fig. 1(b)]. In the 2$\theta$ X-ray diffraction pattern [Fig. 1(c)], only the $(00l)$ peaks are detected, confirming the crystal surface is normal to the $c$ axis. The corresponding electron diffraction pattern [inset in Fig. 1(c)] also confirms the high quality of single crystals.

Figure 1(d) presents the low temperature thermal demagnetization analysis for Fe$_{3-x}$GeTe$_2$ with out-of-plane field using both spin-wave (SW) model and single-particle (SP) model. The temperature dependence of zero-field-cooling (ZFC) magnetization $M(T)$ for Fe$_{3-x}$GeTe$_2$ measured in $H$ = 1 T applied along the $c$ axis is shown in the inset of Fig. 1(d). Localized-moment spin-wave excitations can be described by a Bloch equation:\cite{Das, Kaul1, Kaul2}
\begin{equation}
\frac{\Delta M}{M(0)} = \frac{M(0)-M(T)}{M(0)} = AT^{3/2}+BT^{5/2}+...,
\end{equation}
where $A$ and $B$ are the coefficients. The $M(0)$ is the magnetization at 0 K, which is usually estimated from the extrapolation of $M(T)$. The $T^{3/2}$ term stems from harmonic contribution and the $T^{5/2}$ term is a high-order contribution in spin-wave dispersion relation. In an itinerant magnetism, it is a result of excitation of electrons from one subband to the other. The single-particle excitation is:\cite{Das}
\begin{equation}
\frac{\Delta M}{M(0)} = \frac{M(0)-M(T)}{M(0)} = CT^{3/2}exp{\frac{-\Delta}{k_BT}},
\end{equation}
where $C$, $\Delta$ and $k_B$ are fit coefficient, the energy gap between the Fermi level and the top of the full subband and the Boltzmann constant, respectively. It can be seen that the SW model gives a better fit than the SP model up to 0.9 $T_c$ [Fig. 1(d)], indicating possible localized moment, in agreement with the bad-metallic resistivity of Fe$_{3-x}$GeTe$_2$.\cite{YULIU1} It is also understandable that the SP model fails due to strong electron correlation in Fe$_{3-x}$GeTe$_2$.\cite{Zhu} The fitting yields $A = 8.4(7)\times10^{-5}$ K$^{-3/2}$, $B = 1.24(5)\times10^{-6}$ K$^{-5/2}$, $C = 3.4(1)\times10^{-4}$ K$^{-3/2}$ and $\Delta = 3.9(4)$ meV.

\begin{figure*}
\centerline{\includegraphics[scale=0.85]{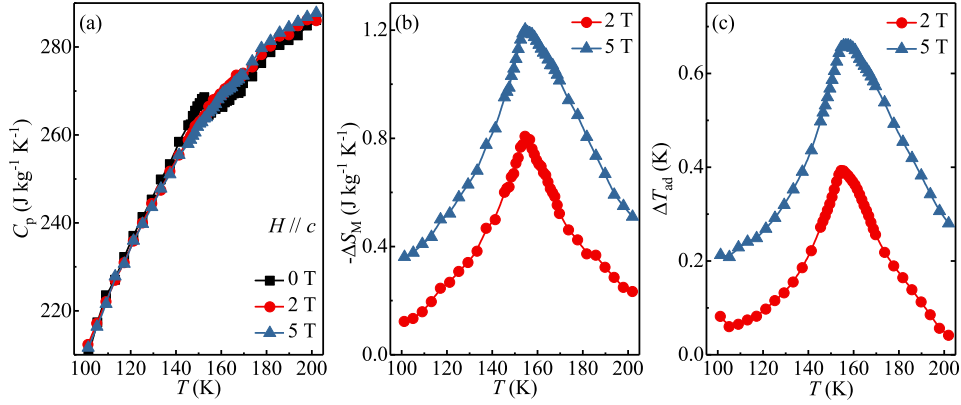}}
\caption{(Color online). Temperature dependences of (a) the specific heat $C_p$, (b) the magnetic entropy change $-\Delta S_M$, and (c) the adiabatic temperature change $\Delta T_{ad}$ for Fe$_{3-x}$GeTe$_2$ with out-of-plane field changes of 2 and 5 T.}
\label{2}
\end{figure*}

Figure 2(a) shows the temperature dependence of heat capacity $C_p$ for Fe$_{3-x}$GeTe$_2$ measured in zero-field and out-of-plane field of 2 and 5 T, respectively. The ferromagnetic order anomaly at $T_c$ = 153 K in the absence of magnetic field is gradually suppressed in fields. The entropy $S(T,H)$ can be determined by
\begin{equation}
S(T,H) = \int_0^T \frac{C_p(T,H)}{T}dT.
\end{equation}
The magnetic entropy change $\Delta S_M(T,H)$ can be approximated as $\Delta S_M(T,H) = S_M(T,H)-S_M(T,0)$. In addition, the adiabatic temperature change $\Delta T_{ad}$ caused by the field change can be derived by $\Delta T_{ad}(T,H) = T(S,H)-T(S,0)$ at constant total entropy $S(T,H)$. Figures 2(b) and 2(c) present the temperature dependence of $-\Delta S_M$ and $\Delta T_{ad}$ estimated from heat capacity with out-of-plane field. They are asymmetric and attain a peak around $T_c$. The maxima of $-\Delta S_M$ and $\Delta T_{ad}$ increase with increasing field and reach the values of 1.20 J kg$^{-1}$ K$^{-1}$ and 0.66 K, respectively, with the field change of 5 T. Since a large magnetic anisotropy is observed in Fe$_{3-x}$GeTe$_2$, it is of interest to further calculate its anisotropic magnetic entropy change.

\begin{figure*}
\centerline{\includegraphics[scale=0.8]{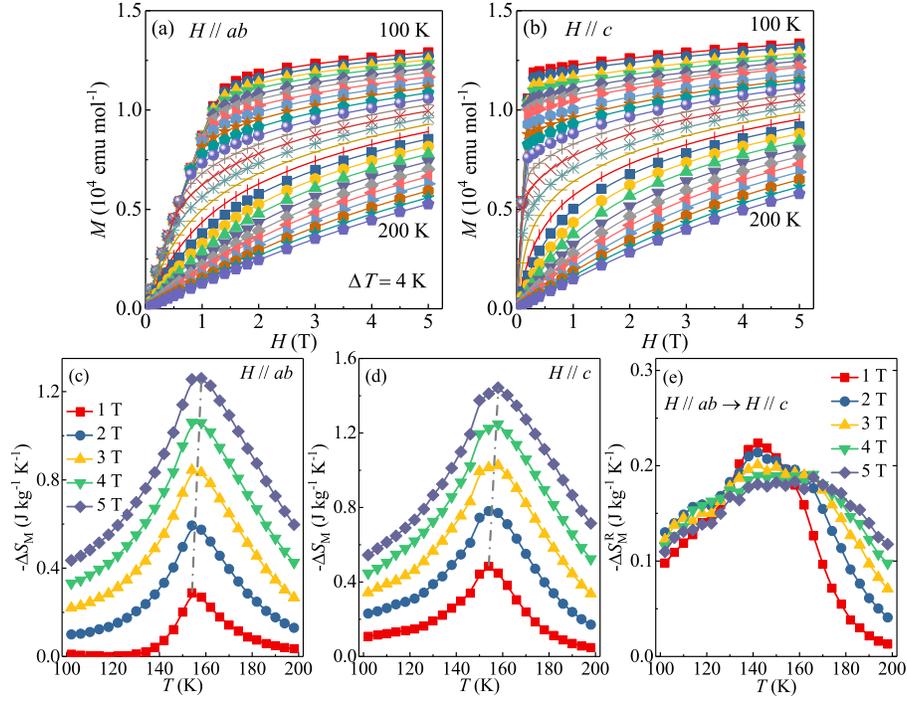}}
\caption{(Color online). Initial isothermal magnetization curves from $T$ = 100 to 200 K with temperature step of $T$ = 4 K measured with (a) in-plane and (b) out-of-plane fields. Temperature dependence of magnetic entropy change $-\Delta S_M$ obtained with (c) in-plane and (d) out-of-plane field changes, and (e) the difference $-\Delta S_M^R$.}
\label{3}
\end{figure*}

Figures 3(a) and 3(b) present the magnetization isotherms with field up to 5 T applied in the $ab$ plane and along the $c$ axis, respectively, in temperature range from 100 to 200 K with a temperature step of 4 K. The magnetic entropy change can be obtained from dc magnetization measurement as:\cite{Pecharsky}
\begin{equation}
\Delta S_M(T,H) = \int_0^H \left[\frac{\partial S(T,H)}{\partial H}\right]_TdH.
\end{equation}
With the Maxwell's relation $\left[\frac{\partial S(T,H)}{\partial H}\right]_T$ = $\left[\frac{\partial M(T,H)}{\partial T}\right]_H$, it can be rewritten as:\cite{Amaral}
\begin{equation}
\Delta S_M(T,H) = \int_0^H \left[\frac{\partial M(T,H)}{\partial T}\right]_HdH.
\end{equation}
When the magnetization is measured at small temperature and field steps, $\Delta S_M(T,H)$ is approximated:
\begin{equation}
\Delta S_M(T,H) = \frac{\int_0^HM(T + \Delta T)dH-\int_0^HM(T)dH}{\Delta T}.
\end{equation}
Figures 3(c) and 3(d) show the calculated $-\Delta S_M(T,H)$ as a function of temperature in various fields up to 5 T applied in the $ab$ plane and along the $c$ axis, respectively. All the $-\Delta S_M(T,H)$ curves feature a pronounced peak around $T_c$, similar to those obtained from heat capacity [Fig. 2(b)], and the peak broadens asymmetrically on both sides with increase in field. Moreover, the value of $-\Delta S_M(T,H)$ increases monotonically with increase in field; the peak $-\Delta S_M$ reaches 1.26 J kg$^{-1}$ K$^{-1}$ with in-plane field change and 1.44 J kg$^{-1}$ K$^{-1}$ with out-of-plane change of 5 T, respectively. We calculated the rotating magnetic entropy change $\Delta S_M^R$ as
\begin{equation}
\Delta S_M^R(T,H) = \Delta S_M(T,H_c)-\Delta S_M(T,H_{ab}).
\end{equation}
The asymmetry of $-\Delta S_M(T,H)$ is more apparent in the temperature dependence of $-\Delta S_M^R$ [Fig. 3(e)]. Furthermore, there is a slight shift of $-\Delta S_M$ maximum towards higher temperature when the field varies from 1 to 5 T [Figs. 3(c) and 3(d)]. This shift of $T_{peak}$ excludes the mean field model but could be reproduced by the Heisenberg model due to its discrepancy with $T_c$.\cite{Francoo}

\begin{figure*}
\centerline{\includegraphics[scale=1.5]{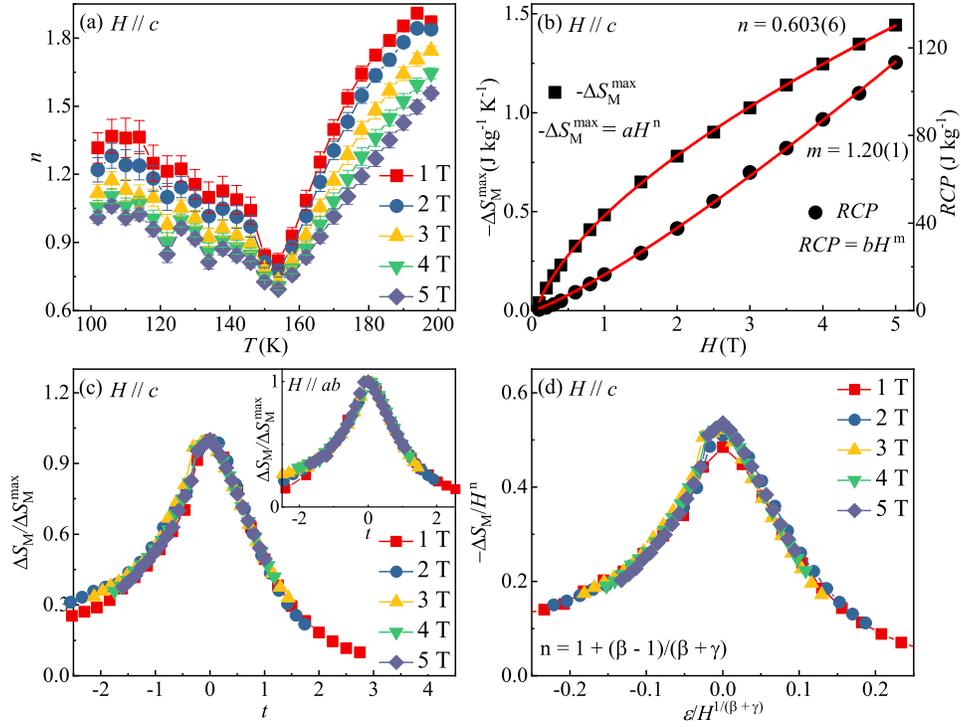}}
\caption{(Color online). (a) Temperature dependence of $n$ in various fields. (b) Field dependence of the maximum magnetic entropy change $-\Delta S_M^{max}$ and the relative cooling power RCP with power law fitting in red solid lines. (c) The normalized $\Delta S_M$ as a function of the rescaled temperature $t$ with out-of-plane field and in-plane field (inset). (d) Scaling plot based on the critical exponents $\beta$ = 0.372 and $\gamma$ = 1.265.\cite{YULIU}}
\label{4}
\end{figure*}

Around the second order phase transition,\cite{Oes} the magnetic entropy maximum change is $-\Delta S_M^{max} = aH^n$,\cite{VFranco} where $a$ is a constant and $n$ is\cite{Francos}
\begin{equation}
n(T,H) = dln\mid \Delta S_M \mid/dln(H).
\end{equation}
Figure 4(a) shows the temperature dependence of $n(T)$ in various fields. All the $n(T)$ curves follow an universal behavior.\cite{FrancoV} At low temperatures, $n$ has a value close to 1. At high temperatures, $n$ tends to 2 as a consequence of the Curie-Weiss law. At $T = T_c$, $n$ has a minimum. Additionally, the exponent $n$ at $T_c$ is related to the critical exponents:\cite{Oes}
\begin{equation}
n(T_c) = 1+\left(\frac{\beta-1}{\beta+\gamma}\right) = 1+\frac{1}{\delta}\left(1-\frac{1}{\beta}\right),
\end{equation}
where $\beta$, $\gamma$, and $\delta$ are the critical exponents related to the spontaneous magnetization $M_s$ below $T_c$, the inverse initial susceptibility $H/M$ above $T_c$, and the isotherm $M(H)$ at $T_c$, respectively.

Relative cooling power (RCP) could be used to estimate the cooling efficiency:\cite{Gschneidner}
\begin{equation}
RCP = -\Delta S_M^{max} \times \delta T_{FWHM},
\end{equation}
where $-\Delta S_M^{max}$ is the entropy change maximum around $T_c$ and $\delta T_{FWHM}$ is the width at half maximum. The RCP also depends on the field as $RCP = bH^m$, where $b$ is a constant and $m$ is related to the critical exponent $\delta$:
\begin{equation}
m = 1+\frac{1}{\delta}.
\end{equation}
Figure 4(b) presents the field-dependent $-\Delta S_M^{max}$ and RCP. The RCP is 113.3 J kg$^{-1}$ within field change of 5 T for Fe$_{3-x}$GeTe$_2$. This is one half of those in manganites and much lower than in ferrites.\cite{Phan, Maalam} Fitting of the $-\Delta S_M^{max}$ and RCP gives $n = 0.603(6)$ and $m = 1.20(1)$, which are close to the values estimated from the critical exponents (Table I).

\begin{table}[ht]
\centering
\begin{tabular}{|l|l|l|l|l|l|}
\hline
Technique & $\beta$ & $\gamma$ & $\delta$ & $n$ & $m$\\
\hline
$-\Delta S_M^{max}$  &   &   &   & 0.603(6) & \\
\hline
RCP  &   &   &   &   & 1.20(1)\\
\hline
MAP & 0.374(1) & 1.273(8) & 4.404(12) & 0.620(1) & 1.227(1) \\
\hline
KFP & 0.372(4) & 1.265(15) & 4.401(6) & 0.616(2) & 1.227(1) \\
\hline
CI  &   &   & 4.50(1) & & 1.222(1) \\
\hline
\end{tabular}
\caption{\label{tab}Critical exponents of Fe$_{3-x}$GeTe$_2$.\cite{YULIU} The MAP, KFP and CI represent the modified Arrott plot, the Kouvel-Fisher plot and the critical isotherm, respectively.}
\end{table}

The scaling of magnetocaloric data is constructed by normalizing all the $-\Delta S_M$ curves against the maximum $-\Delta S_M^{max}$, namely, $\Delta S_M/\Delta S_M^{max}$ by rescaling the temperature $t$ below and above $T_c$ as defined in:
\begin{equation}
t_- = (T_{peak}-T)/(T_{r1}-T_{peak}), T<T_{peak},
\end{equation}
\begin{equation}
t_+ = (T-T_{peak})/(T_{r2}-T_{peak}), T>T_{peak},
\end{equation}
where $T_{r1}$ and $T_{r2}$ are the temperatures of two reference points corresponding to $\Delta S_M(T_{r1},T_{r2}) = \frac{1}{2}\Delta S_M^{max}$.\cite{Franco} All the $-\Delta S_M(T,H)$ curves collapse onto a single curve regardless of temperature and field, as shown in Fig. 4(c). In the phase transition region, the scaling analysis of $-\Delta S_M$ can also be expressed as
\begin{equation}
\frac{-\Delta S_M}{a_M} = H^nf(\frac{\varepsilon}{H^{1/\Delta}}),
\end{equation}
where $a_M = T_c^{-1}A^{\delta+1}B$ with A and B representing the critical amplitudes as in $M_s(T) = A(-\varepsilon)^\beta$ and $H = BM^\delta$, $\Delta = \beta + \gamma$, and $f(x)$ is the scaling function.\cite{Su} If the critical exponents are appropriately chosen, the $-\Delta S_M(T)$ curves should be rescaled into a single curve, consistent with normalizing all the $-\Delta S_M$ curves with two reference temperatures. By using the values of $\beta$ = 0.372 and $\gamma$ = 1.265 obtained by the Kouvel-Fisher plot,\cite{YULIU} we have replotted the scaled $-\Delta S_M$ for Fe$_{3-x}$GeTe$_2$ [Fig. 4(d)]. The good overlap of the experimental data points clearly indicates that the obtained critical exponents for Fe$_{3-x}$GeTe$_2$ are not only in agreement with the scaling hypothesis but also intrinsic.

\begin{figure*}
\centerline{\includegraphics[scale=0.8]{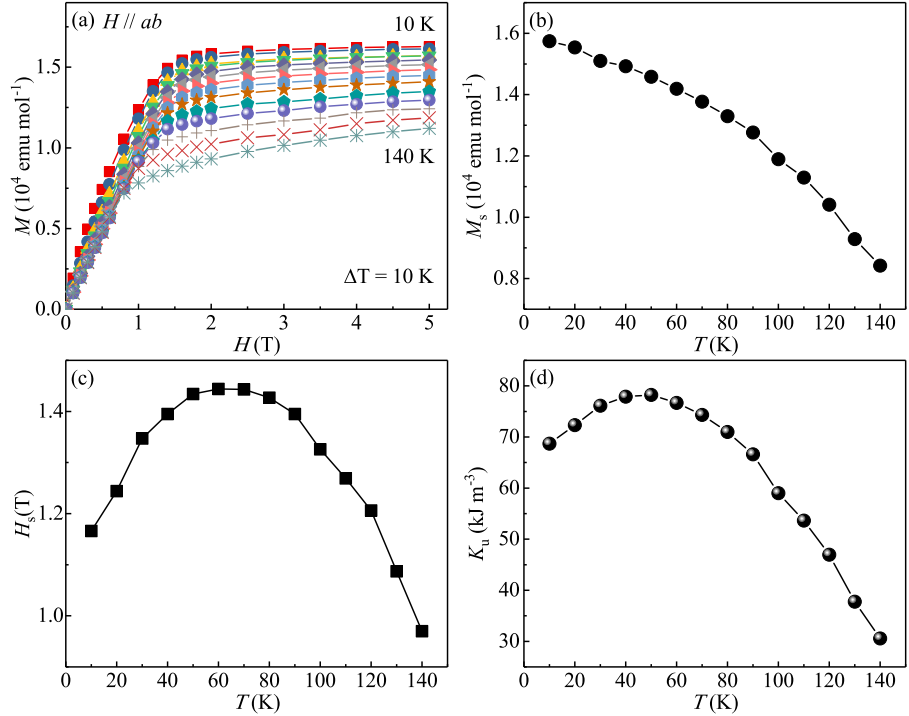}}
\caption{(Color online). (a) Initial isothermal magnetization curves from $T$ = 10 to 140 K with in-plane fields. Temperature evolution of (b) the saturation magnetization $M_s$, (c) the saturation field $H_s$, and (d) the anisotropy constant $K_u$.}
\label{5}
\end{figure*}

Then we estimated the magnetocrystalline anisotropy of Fe$_{3-x}$GeTe$_2$. By using the Stoner-Wolfarth model a value for the magnetocrystalline anisotropy constant $K_u$ can be estimated from the saturation regime in isothermal magnetization curves [Fig. 5(a)].\cite{Stoner} Within this model the magnetocrystalline anisotropy in the single domain state is related to the saturation magnetic field $H_s$ and the saturation moment $M_s$ with $\mu_0$ is the vacuum permeability:
\begin{equation}
\frac{2K_u}{M_s} = \mu_0H_{sat}.
\end{equation}
When $H // ab$, the anisotropy becomes maximal. We estimated the saturation magnetization $M_s$ by using a linear fit of $M(H)$ above a magnetic field of 2.5 T with in-plane field [Fig. 5(b)], which monotonically decreases with increasing temperature. Then we determined the saturation field $H_s$ as the intersection point of two-linear fits, one being a fit to the saturated regime at high fields and one being a fit of the unsaturated linear regime at low fields. The value of $H_s$ increases at low temperature, which is possibly related to a spin reorientation transition,\cite{YULIU1} and then decreases with increasing temperature [Fig. 5(c)]. Figure 5(d) presents the temperature evolution of $K_u$ for Fe$_{3-x}$GeTe$_2$, which can not be described by the $l(l+1)/2$ power law.\cite{Callen,Mryasov} The value of $K_u$ for Fe$_{3-x}$GeTe$_2$ is about 69 kJ cm$^{-3}$ at 10 K, slightly increases to 78 kJ cm$^{-3}$ at 50 K, and then decrease with increasing temperature, which are comparable to those for CrBr$_3$, but smaller than those for CrI$_3$.\cite{Richter} The decrease of $K_u$ with increasing temperature is also observed in CrBr$_3$ and CrI$_3$,\cite{Richter} arising from a large number of local spin clusters.\cite{Zener,Carr} In a pure two-dimensional system, materials with isotropic short-range exchange interactions can not magnetically order. The long-range ferromagnetism in few-layers of Fe$_{3-x}$GeTe$_2$ could possibly be favored by the large magnetocrystalline anisotropy.

\section*{Conclusion}

In summary, we have investigated in detail the magnetocaloric effect of Fe$_{3-x}$GeTe$_2$ single crystals. The large magnetocrystalline anisotropy is found to be temperature-dependent and probably establishes the long-range ferromagnetism in few-layers of Fe$_{3-x}$GeTe$_2$. The magnetic entropy change $-\Delta S_M$ also reveals an anisotropic characteristic and could be well scaled into a universal curve independent on temperature and field. By fitting of the field-dependent parameters of $-\Delta S_M^{max}$ and the relative cooling power RCP, it gives $-\Delta S_M^{max} \propto H^n$ with $n = 0.603(6)$ and $RCP \propto H^m$ with $m = 1.20(1)$ when $H // c$. Considering its tunable room-temperature ferromagnetism and hard magnetic properties in nanoflakes, further investigation on the size dependence of magnetocaloric effect is of interest.

%\clearpage
%\newpage
%\bibliography{all}

\section*{Acknowledgements}

We thank J. Warren for help with the scanning electron microscopy (SEM) measurement. Work at Brookhaven National Laboratory is supported by the US DOE, Contract No. DE-SC0012704.

\section*{Author contributions statement}

Y.L. and C.P. designed this study and synthesized crystals; Y. L. performed magnetization and heat capacity measurements. J.L, J.T. and Y.Z. contributed TEM measurement. Y.L. and C.P. organized and wrote the paper with input from all collaborators. This manuscript reflects the contribution and ideas of all authors.

\section*{Additional information}
Competing interests: The authors declare no competing interests.

\end{document}